# CHEMICAL COMPOSITION OF METAL-POOR STARS IN COMA BERENICES ULTRA-FAINT DWARF GALAXY AS A PROXY TO INDIVIDUAL CHEMICAL ENRICHMENT EVENTS

**Sitnova T.M.[1], Mashonkina L.I.[1], Tatarnikov A.M.[2], Voziakova O.V.[2], Burlak M.A.[2], Pakhomov Yu.V[1]**

[1] *INASAN, Moscow, Russia*
[2] *SAI MSU, Moscow, Russia*
*sitnova@inasan.ru*

*We present NLTE abundances and atmospheric parameters for three metal-poor stars ([Fe/H] < -2) in Coma Berenices ultra-faint dwarf galaxy (UFD). The derived results are based on new photometric observations in the visible and infra-red bands obtained with the 2.5-m telescope of the SAI MSU Caucasian observatory and high-resolution spectra from the archive of the 10-m Keck telescope. Effective temperatures ($T_{eff}$) were determined from V-I, V-K, V-J colours and colour-$T_{eff}$ relation. For each star, temperatures derived from different colours agree within 20 K. Surface gravities (log g) were calculated using a well-known relation between log g, $M_V$, bolometric correction, distance (d = 44 kpc), $T_{eff}$, and mass, adopted as 0.8 $M_{Sun}$. The NLTE abundances for Na, Mg, Ca, Ti, Fe, Ni, Sr, and Ba were determined. Our results differ from those available in the literature, derived under the LTE assumption. A revision of atmospheric parameters and abundances, based on new photometric observations and accurate modelling of spectral line formation resulted in reinterpretation of the star formation history in Coma Berenices UFD. The derived chemical abundance patterns for three stars differ from each other and from that, which is typical for the MW halo stars of similar [Fe/H]. The S1 star shows solar [α/Fe] and unprecedentedly low [Na/Mg] of -1.46, which is the lowest value among all metal-poor stars known to date. Abundance pattern of the S1 star is well reproduced by nucleosynthesis model in metal-free massive stars explosion. The stars S2 and S3, in contrast to S1, show high [α/Fe] ratios, for example, [Mg/Fe] = 0.8 in S2, while stars with -3.5 < [Fe/H] < -2 in the other dwarf galaxies and the MW halo show a typical ratio [Ca,Mg,Ti/Fe] of 0.3 dex. All the three stars show low Sr and Ba abundances. These peculiarities in chemical composition argue for a small number of nucleosynthesis events contributed to chemical abundances of these stars. A wide enough range of metallicity, 0.65 dex, observed in Coma Berenices is likely produced by inhomogeneous mixing of the interstellar medium, but not an increase in [Fe/H] during a prolonged star formation.*

One of the smallest known to date ultra-faint dwarf (UFD) galaxies in the Local Group is a Coma Berenices UFD with $M_V$ = -3.7, half-light radius $r_{1/2}$ = 70 pc [1] and stellar mass of 3700 $M_{Sun}$ [2]. From analysis of colour-magnitude diagram, Belokurov et al. [1] found that stars of the galaxy form a homogeneous population with an age of 12 Gyr and a metallicity of [Fe/H] = -2. Using high-resolution spectra for three brightest stars with V ~ 18 mag, Frebel et al. [3] found that the metallicity spans in the -2.88 < [Fe/H] < -2.31 range and concluded that more observations are required to understand whether a wide [Fe/H] range is caused by inhomogeneous mixing in the interstellar medium or by gradual metallicity increase during a prolonged star formation. Vargas et al. [4, V13] increased the statistics and determined abundances of Mg, Ca, Si, Ti, and Fe in 10 stars using medium resolution spectra. V13 stellar sample shows metallicities in the -3.38 < [Fe/H] < -2.12 range and different [α/Fe] ratios: low [α/Fe] ~ 0 in two stars with [Fe/H] > -2.4 and high [α/Fe] ~ 0.3 to 0.9 dex in eight stars with [Fe/H] < -2.4. V13 concluded that low [α/Fe] ratio in two stars with the highest [Fe/H] is caused by production of iron started in SN Ia. It is worth noting that the prolonged star formation in such a small system as Coma Berenices is unlikely and contradicts to results of analyses of colour-magnitude diagram [1]. Moreover, a peculiar abundance pattern found in ComaBer S-1 [3] and, particularly, an extremely low sodium abundance of [Na/Fe] = -0.74, cannot be explained with nucleosynthesis in SN Ia. V13 did not determine sodium abundance, since their observed spectral range does not cover any sodium lines.

This study is conducted within a project of a systematic study of chemical composition of stars in dwarf galaxies using a common method of atmospheric parameters and abundances determination, as described in our earlier paper [5], where we studied metal-poor, [Fe/H] < -1.5, stars in seven dSph galaxies and the Milky Way.

Effective temperatures ($T_{eff}$) were determined from V-I, V-K, V-J colours and colour-$T_{eff}$ relation. Photometric observations were carried out with 2.5-m telescope of CMO of SAI MSU at nights with stable extinction, which was monitored by astroclimate site monitor of CMO [6]. Observations in BVRcIc bands were carried out on 16 and 19 April, 2017 using an NBI 4k x 4k camera produced by N. Bohr Institute, Copenhagen. A procedure of preliminary reduction of frames contains bias subtraction, nonlinearity and flat field correction. Magnitudes were derived with sextractor package [7] taking into account aperture corrections. Coordinates of the objects were derived with astrometry package [8]. Extinction measurements and photometric calibration were made using Landolt standards [9], observed at the same nights. The equations of the transformation to the standard photometric system were derived earlier using Landolt standards [10]. The JHK photometric data were derived on 18 February and 27 April, 2017 with ASTRONIRCAM camera [11] at the dithering mode. During each night, 30 frames for each filter were derived with a total exposure time of 900 seconds. Nonlinearity, dark current, and flat field corrections of frames were performed. Photometry was carried out in a MKO system, with a further transformation to 2MASS according to equations from [12]. As the main standard, we adopted the star GSPC P264-F from the list of standards of this photometric system [12], to which bright enough stars in the field were referred and served as reference stars for every investigated object. The derived magnitudes are listed in Table 1. An error of each magnitude does not exceed several hundredths, and, for each star, temperatures derived from different colours agree within 20 K. Errors of 2MASS catalogue for the investigated objects are significantly larger. For example, for S1 star, K = 16.013 ± 0.286, which leads to an uncertainty of 330 K in $T_{eff}$ from V-K colour.

Table 1.
Photometric magnitudes derived with the 2.5-m telescope of CMO of SAI MSU for three stars in the Coma Berenices UFD.

| Star | B | V | I | R | J | H | K |
|---|---|---|---|---|---|---|---|
| ComaBer S1 | 18.920 | 18.164 | 17.678 | 17.197 | 16.514 | 15.996 | 15.955 |
| ComaBer S2 | 18.375 | 17.572 | 17.058 | 16.552 | 16.407 | 15.885 | 15.834 |
| ComaBer S3 | 18.854 | 18.099 | 17.608 | 17.115 | 15.818 | 15.272 | 15.213 |

Surface gravities (log g) were calculated using a well-known relation between log g, $M_V$, bolometric correction, distance (d = 44 kpc [1]), $T_{eff}$, and mass, adopted as 0.8 $M_{Sun}$. For stars S1, S2, and S3 we derived $T_{eff}$ / log g / [Fe/H] = 4900 / 2.0 / -2.08, 4875 / 1.95 / -2.73, and 4785 / 1.70 / -2.46. For the same stars, Frebel et al. [3] derived 4700 / 1.3 / -2.31, 4600 / 1.4 / -2.88, and 4600 / 1.0 / -2.53, respectively, from LTE analysis of Fe I and Fe II lines.

High-resolution observed spectra (R = 34000) were taken from the archive of the Keck telescope. The NLTE abundances for Na, Mg, Ca, Ti, Fe, Ni, Sr, and Ba were determined. The derived results for three stars are plotted in Fig. 1 together with the abundances of a halo star HD 122563, which has similar atmospheric parameters (4600 / 1.3 / -2.6), and the calculated yields of SN II explosions from [13]. The derived abundance patterns for the three stars differ from each other and from those for the halo star. The star S1 shows extremely low sodium abundance and solar [α/Fe] ratio. The stars S2 and S3, in contrast to S1, show high [α/Fe] ratios, for example, [Mg/Fe] = 0.8 in S2, while stars with -3.5 < [Fe/H] < -2 in the other dwarf galaxies and the MW halo show a typical ratio [Ca,Mg,Ti/Fe] of 0.3 dex. All the three stars show low Sr and Ba abundances. In the MW halo stars, peculiar chemical abundance patterns are mostly found at [Fe/H] < -4 and are explained with individual nucleosynthesis events and inhomogeneous mixing of the interstellar medium [13]. We tried to fit the derived abundances for elements from Na to Ni with the predictions of nucleosynthesis in the explosions of massive metal-free stars, as calculated by [13]. When using ratios [El/Mg] = $\log(N_{el}/N_{Mg})_{Star}$ - $\log(N_{el}/N_{Mg})_{Sun}$, peculiar chemical abundance pattern of the S1 star is well reproduced by model with $M_{init}$ = 80 $M_{Sun}$, $E_{expl}$ = 10 B, and $f_{mix}$ = 0.1. However, in order to fit the S1 pattern in the absolute abundance scale, [El/H], no less than four explosions are required. The derived abundance patterns of S2 and S3 cannot be explained with one of the models available in the grid [13]. However, large [α/Fe] overabundances argue for a small number of nucleosynthesis events contributed to chemical abundances of these stars.

A revision of atmospheric parameters and abundances, based on new photometric observations and accurate modelling of spectral line formation resulted in reinterpretation of the star formation history in Coma Berenices UFD galaxy. The derived chemical abundance patterns for three stars differ from each other and from that, which is typical for the MW halo stars of similar [Fe/H]. This argues for a small number of nucleosynthesis events contributed to chemical abundances of these stars. A wide enough range of metallicity observed in Coma Berenices is likely produced by inhomogeneous mixing of the interstellar medium, but not an increase in [Fe/H] during a prolonged star formation.

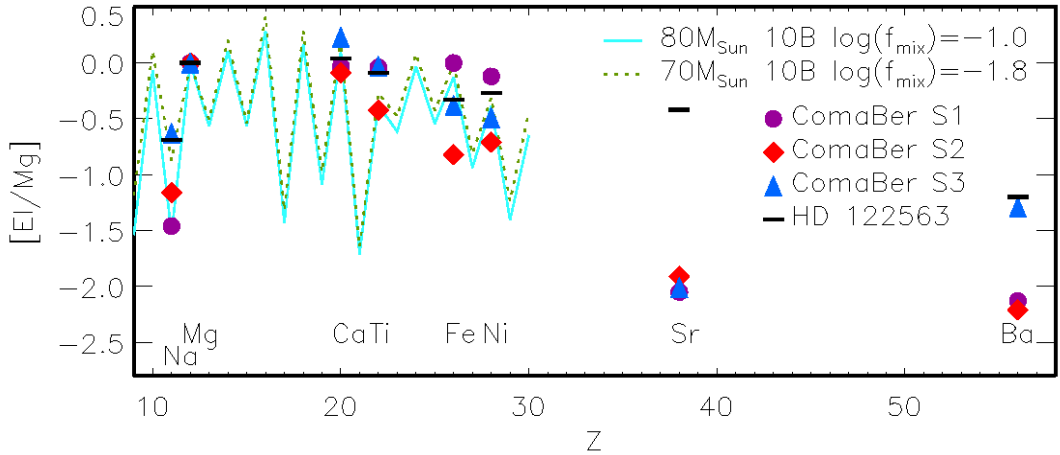

Fig. 1. The abundance ratios relatively to magnesium in three Coma Berenices stars and HD 122563 as a function of an atomic number. The nucleosynthesis calculations are from [13].

This work was supported in part by M.V.Lomonosov Moscow State University Program of Development. We made use of the Keck Observatory Archive (C168Hr) and the StarFit database http://starfit.org/. L.M., T.S., and Y.P. are grateful to Program of Presidium of the RAS, P-28 for financial support.